\makeatletter
\def\input@path{{tex/}{bst/}}
\makeatother

\documentclass[%
 reprint,
 amsmath,amssymb,
 aps,
]{revtex4-2}

\usepackage{graphicx}
\usepackage[dvipsnames]{xcolor}
\usepackage{lipsum}
\usepackage{dcolumn}
\usepackage{bm}
\usepackage{hyperref}

\usepackage{xspace}

\newcommand{\figref}[2][]{\autoref{#2}#1\xspace}

\newcommand{\Eqref}[2][]{Eq.~\ref{#2}#1\xspace}

\newcommand{\Nsweeps}{\ensuremath{N_\mathrm{sweeps}}\xspace}
\newcommand{\Nreps}{\ensuremath{N_\mathrm{reps}}\xspace}
\newcommand{\Nsteps}{\ensuremath{N_\mathrm{steps}}\xspace}

\newcommand{\WSC}{WSC\xspace}

\newcommand{\RL}{RL\xspace}

\newcommand{\numTestRuns}{100\xspace}
\newcommand{\numTestInstances}{100\xspace}
\newcommand{\numNsteps}{40\xspace}
\newcommand{\numNsweeps}{100\xspace}
\newcommand{\numEps}{25, 000\xspace}
\newcommand{\numNreps}{64\xspace}
\newcommand{\numPPOGamma}{0.99\xspace}
\newcommand{\numLearningRate}{\ensuremath{1\times 10^{-6}}\xspace}

\renewcommand{\omit}[1]{}
\newcommand{\added}[1]{#1} 
\newcommand{\loss}[1]{L^{\text{#1}}(\theta)}

\begin{document}

\preprint{}

\title{Finding the ground state of spin Hamiltonians with reinforcement learning}

\author{Kyle Mills}
\email{kyle.mills@1qbit.com}
\affiliation{%
 1QB Information Technologies (1QBit), Vancouver, British Columbia, Canada\\
 University of Ontario Institute of Technology, Oshawa, Ontario, Canada\\
 Vector Institute for Artificial Intelligence, Toronto, Ontario, Canada}%

\author{Pooya Ronagh}%
\email{pooya.ronagh@1qbit.com}
\affiliation{%
  1QB Information Technologies (1QBit), Vancouver, British Columbia, Canada\\
  Institute for Quantum Computing (IQC), Waterloo, Ontario, Canada\\
  Department of Physics and Astronomy, University of Waterloo, Ontario, Canada
}%

\author{Isaac Tamblyn}%
\email{isaac.tamblyn@nrc.ca}
\affiliation{%
 National Research Council Canada, Ottawa, Ontario, Canada\\
 University of Ottawa, Ottawa, Ontario, Canada\\
 Vector Institute for Artificial Intelligence, Toronto, Ontario, Canada
}%

\date{\today}

\begin{abstract}
Reinforcement learning (RL) has become a proven method for optimizing a procedure for which success has been defined, but the specific actions
needed to achieve it have not.  Using a method we call ``Controlled Online Optimization Learning'' (COOL), we apply the so-called ``black box'' method of
RL to  simulated annealing (SA),
demonstrating that an RL agent based on proximal policy optimization can, through
experience alone, arrive at a temperature schedule that surpasses the
performance of standard heuristic temperature schedules for two classes of
Hamiltonians.  When the system is initialized at a cool temperature, the RL
agent learns to heat the system to ``melt'' it, and then slowly cool it in an effort to anneal
to the ground state; if the system is initialized at a high temperature, the
algorithm immediately cools the system.  We investigate the performance of our
RL-driven SA agent in generalizing to all Hamiltonians of a specific
class; when trained on random Hamiltonians of nearest-neighbour spin glasses,
the RL agent is able to control the SA process for other
Hamiltonians, reaching the ground state with a higher probability than a simple
linear annealing schedule. Furthermore, the scaling performance (with respect to
system size) of the RL approach is far more favourable, achieving a performance improvement of almost two orders of
magnitude on $L=14^2$ systems.  We demonstrate the robustness
of the RL approach when the system operates in a ``destructive observation''
mode, an allusion to a quantum system where measurements destroy the state of
the system. The success of the RL agent could have far-reaching impact, from
classical optimization, to quantum annealing, to the simulation of physical
systems.
\end{abstract}

\maketitle


\section{Introduction}

In metallurgy and materials science, the process of annealing is used to
equilibrate the positions of atoms to obtain perfect low-energy crystals. Heat
provides the energy necessary to break atomic bonds, and high-stress interfaces
are eliminated by the migration of defects. By slowly cooling the metal to room
temperature, the metal atoms become energetically locked in a lattice structure
more favourable than the original structure. Metallurgists can tune the
temperature schedule to arrive at final products that have desired characteristics,
such as ductility and hardness. Annealing is a biased stochastic search for the
ground state.

\begin{figure}
    \centering
    \includegraphics[width=\columnwidth]{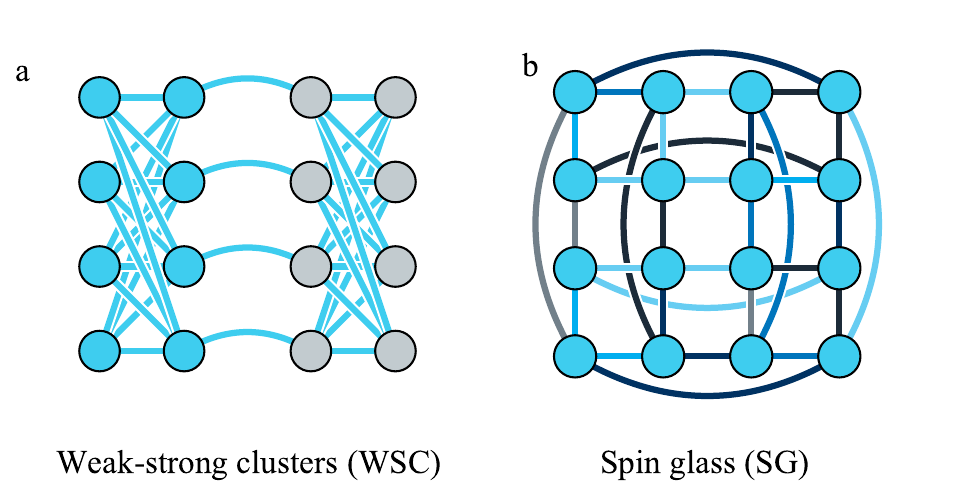}
    \includegraphics[width=\columnwidth]{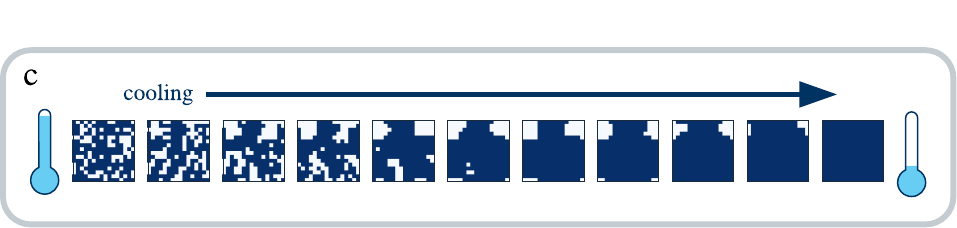}
    \caption{
    Two classes of Hamiltonian problems are depicted. (a) The weak-strong
    clusters (WSC) model comprises two bipartite clusters.  The left
    cluster is biased upward ($h_i>0$); the right cluster is biased downward ($h_i < 0$). All
    couplings are equal and of unit magnitude. The two clusters are coupled
     via the eight central nodes.  This model exhibits a deep local
    minimum very close in energy to the model's global minimum. When initialized
    in the local minimum, the \RL agent is able to learn schemes to escape
    the local minimum and arrive at the global minimum, without any explicit
    knowledge of the Hamiltonian.  
    (b) Here we present an example spin-glass model.
    The nodes are coupled to nearest neighbours with random Gaussian-distributed
    coupling coefficients.  The nodes are unbiased ($h_i=0$), and the couplings are
    changed at each instantiation of the model. The \RL algorithm is able to learn
    a dynamic temperature schedule by observing the system throughout the
    annealing process, without explicit knowledge of the form of the
    Hamiltonian, and the learned policy can be applied to all instances of
    randomly generated couplings. We demonstrate this on variably sized spin glasses and investigate the scaling with respect to a classic linear
    SA schedule.  In (c), we show snapshots of a sample progression of a configuration undergoing SA under the ferromagnetic Ising model Hamiltonian and a constant cooling schedule. The terminal state, all spins-up, is the ground state; this anneal would be considered successful.}
    \label{fig:hamiltonians}
\end{figure}

An analogous \textit{in silico} technique, simulated annealing (SA)~\cite{Kirkpatrick1983a}, can be used to find the ground state of spin-glass
models, an NP-hard problem \cite{Barahona1982}.  A spin glass is a graphical
model consisting of binary spins $\sigma_i$.  The connections between spins are
defined by the coupling constants $J_{ij}$, and a linear term with coefficients
$h_i$ can apply a bias to individual spins. The Hamiltonian
\begin{equation*}
     \mathcal H = -\sum_{i< j}J_{ij}\sigma_i\sigma_j - \sum_i h_i\sigma_i,\quad \sigma_i=\pm 1
\end{equation*}
defines the energy of the microstates \cite{Sherrington1975}.  The choices of
the quadratic coupling coefficients $J_{ij}$ and the linear bias coefficients
$h_i$ effect the interesting dynamics of the model: $J_{ij}$ can be randomly
distributed according to a Gaussian distribution  \cite{Sherrington1975},
encompass all $i,j$ combinations for a fully connected Hamiltonian, or be
limited to short-range (e.g., nearest-neighbour, $\langle i, j \rangle$)
interactions, to name a few. For example, when the positive, unit-magnitude
coupling is limited to nearest-neighbour pairs, the ubiquitous ferromagnetic
Ising model \cite{Ising1925} is recovered. Examples of the Hamiltonians we
investigate in this work are presented in \figref{fig:hamiltonians} and
discussed in further detail in Section~\ref{sec:Hamiltonians}.

Finding the ground state of (i.e., ``solving'') such systems is interesting from the perspective 
of thermodynamics, as one can observe phenomena such as phase
transitions \cite{Onsager1944,Ferdinand1969}, but also practically useful as
discrete optimization problems can be mapped to spin-glass models (e.g., the
travelling salesperson problem or the knapsack problem) \cite{Lucas2014}. The
Metropolis--Hastings algorithm \cite{Hastings1970,Metropolis1953} can be used to
simulate the spin glass at arbitrary temperature, $T$; thus, it is used ubiquitously
for SA. By beginning the simulation at a high temperature, one
can slowly cool the system over time, providing sufficient thermal energy to
escape local minima, and arrive at the ground state ``solution'' to the problem.
The challenge is to find a temperature schedule that minimizes computational
effort while still arriving at a satisfactory solution; if the temperature is
reduced too rapidly, the system will become trapped in a local minimum, and reducing the temperature  too slowly results in an unnecessary computational expense. Kirkpatrick et al.
\cite{Kirkpatrick1983a,Kirkpatrick1984} proposed starting at a temperature that
results in an 80\% acceptance ratio (i.e., 80\% of Metropolis spin flips are
accepted) and reducing the temperature geometrically.  They also recommended
monitoring the objective function and reducing the cooling rate if
the objective value (e.g., the energy) drops too quickly.  More-sophisticated adaptive temperature
schedules have been investigated~\cite{VanLaarhoven1987}\added{; however, simple linear and reciprocal temperature schedules are commonly used in practice \cite{stander1994temperature,Heim215}. We will refer to SA using a linear schedule as ``classic SA'' throughout this work. } Nevertheless, in his
1987 paper, Bounds \cite{Bounds1987} said that ``choosing an annealing schedule
for practical purposes is still something of a black art''.

When framed in the advent of quantum computation and quantum control,
establishing robust and dynamic scheduling of control parameters becomes even
more relevant. For example, the same optimization problems that can be cast as
classical spin glasses are also amenable to quantum annealing
\cite{Farhi2001,Hen2012,Boixo2014,Bian2014,Venturelli2015}, exploiting, in lieu
of thermal fluctuations, the phenomenon of quantum tunnelling~\mbox{\cite{Ray1989,Martonak2002,Santoro2002}} to escape local minima. Quantum
annealing (QA) was proposed by Finnila et al. \cite{Finnila1994} and Kadowaki
and Nishimori \cite{Kadowaki1998}, and, in recent years, physical realizations of
devices capable of performing QA (quantum annealers), have been
developed \cite{Harris2010,Harris2010a,Johnson2011,McGeoch2013}, and are being
rapidly commercialized.  As these technologies progress and become more
commercially viable, practical applications \cite{Ikeda2019,Venturelli2015} will
continue to be identified and resource scarcity will spur the already extant
discussion of the efficient use of annealing hardware
\cite{Dickson2013,Okada2019}.

Nonetheless, there are still instances where the classical (SA) outperforms the
quantum (QA) \cite{Battaglia2005}, and improving the former should not be
undervalued. \textit{In silico} and hardware annealing solutions such as Fujitsu's
FPGA-based Digital Annealer \cite{Tsukamoto2017}, NTT's laser-pumped coherent
Ising machine (CIM) \cite{Inagaki2016,Leleu2019,Tiunov2019}, and the quantum circuit model algorithm
known as QAOA \cite{Farhi2014,Farhi2019} all demand the scheduling of control
parameters, whether it is the temperature in the case of the Digital Annealer, or the power
of the laser pump in the case of  CIM. Heuristic methods based on trial-and-error
experiments are commonly used to schedule these control parameters, and an
automatic approach could expedite  development, and improve the stability of
such techniques.

In this work, we demonstrate the use of a reinforcement learning (RL) method to learn
the ``black art'' of SA temperature scheduling, and show
that an RL agent is able to learn dynamic control parameter
schedules for various problem Hamiltonians. The schedules that the RL agent produces are dynamic and reactive, adjusting to the current
observations of the system to reach the ground state quickly and consistently
without \textit{a priori} knowledge of a given Hamiltonian.  \added{We believe that RL will be important for quantum information processing, especially for hardware- and software-based control.}

\begin{figure*}
    \centering
    \includegraphics[width=\textwidth]{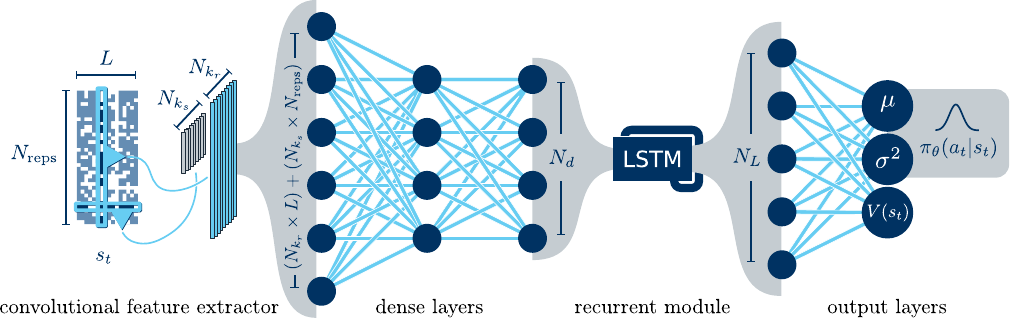}
    \caption{
    A neural network is used to learn the control parameters for several
    SA experiments. By observing a lattice of spins, the neural
    network can learn to control the temperature of the system in a
    dynamic fashion, annealing the system to the ground state. The spins at time
    $t$ form the state $s_t$ fed into the network. Two concurrent convolutional
    layers extract features from the state. These features are combined with a
    dense layer and fed into a recurrent module (an LSTM module) capable of
    capturing temporal characteristics. The LSTM module output is reduced to two
    parameters used to form the policy distribution $\pi_\theta(a_t\mid s_t)$ as
    well as to approximate the value function $V(s_t)$ used for the generalized
    advantage estimate.  }
    \label{fig:schematic}
\end{figure*}

\section{The environment and architecture}

\subsection{Reinforcement learning}

Reinforcement learning is a branch of dynamic programming whereby an agent,
residing in state $s_t$ at time $t$, learns to take an action $a_t$ that
maximizes a cumulative reward signal $R$ by dynamically interacting with an
environment \cite{Sutton2018}. Through the training process, the agent arrives
at a policy $\pi$ that depends on some observation (or ``state'') of the system,
$s$.  In recent years, neural networks have taken over as the \textit{de facto}
function approximator for the policy. Deep reinforcement learning has seen
unprecedented success, achieving superhuman performance in a variety of video
games \cite{Berner2019,Zhang2019,Vinyals2019,Mnih2013}, board games
\cite{Silver2016,Silver2017,Silver2018}, and other puzzles
\cite{Agostinelli2019,OpenAI2019}.  While many reinforcement learning algorithms
exist, we have chosen to use proximal policy optimization (PPO)
\cite{Schulman2017}, implemented within Stable Baselines \cite{stable-baselines}
for its competitive performance on problems with continuous action spaces.

\subsection{The environment}

We developed an OpenAI gym \cite{Brockman2016} environment which serves as
the interface to the ``game'' of simulated annealing.  Let us now define some
terminology and parameters important to SA.  For
a given Hamiltonian, defining the interactions of $L$ spins, we create \Nreps
randomly initialized replicas (unless otherwise specified). The initial spins of
each replica are drawn from a Bernoulli distribution with probability of a spin-up
being randomly drawn from a uniform distribution.  These independent replicas
are annealed in parallel. The replicas follow an identical temperature schedule
with their uncoupled nature providing a mechanism for statistics of the system to
be represented through an ensemble of measurements.  In the context of the
Metropolis--Hastings framework, we define one ``sweep'' to be $L$ \added{proposed} random spin
flips (per replica), and one ``step'' to be \Nsweeps. After every step, the
environment returns an observation of the current state $s_t$ of the system, an
$\Nreps\times L$ array consisting of the binary spin values present. This
observation can be used to make an informed decision of the action $a_t$ that
should be taken.  The action, a single scalar value, corresponds to the total
inverse temperature change $\Delta\beta$ \added{(where $\beta=1/T$)} that should be carried out over the
subsequent step. The choice of action is provided to the environment, and the
process repeats until \Nsteps steps have been taken, comprising one full anneal,
or ``episode'' in the language of \RL. If the chosen action would result in the
temperature becoming negative, no change is made to the temperature and the
system continues to evolve under the previous temperature.

\added{In our investigations, we choose $\Nsteps=40$ and  $\Nsweeps=100$ resulting, in 4000 sweeps per episode.  These values define the maximum size of system we can compare to classic SA. This number of sweeps is sufficient for a linear schedule to attain measurable success on all but the largest system size we investigate.}

\subsection{Observations}

For the classical version of the problem, an observation consists of the explicit spins of an
ensemble of replicas. In the case of an unknown Hamiltonian, the ensemble
measurement is important as the instantaneous state of a single replica does not
provide sufficient information about the current temperature of the system.
Providing the agent with multiple replicas allows it to compute statistics and
have the possibility of inferring the temperature. For example, if there is
considerable variation among replicas, then the system is likely hot, whereas if
most replicas look the same, the system is probably cool.

When discussing a quantum system, where the spins represent qubits, direct
mid-anneal measurement of the system is not possible as measurement causes a
collapse of the wavefunction. To address this, we discuss
experiments conducted in a ``destructive observation'' environment, where
measurement of the spins is treated as a ``one-time'' opportunity for inclusion
in \RL training data.  The subsequent observation is then based on a different
set of replicas that have evolved through the same schedule, but from different
initializations.

\added{When running the classic SA baselines, to keep comparison fair, each episode consists of \Nreps replicas as in the \RL case. If even one replica reaches the ground state, the episode is considered a success.}

\subsection{Reinforcement learning algorithm}

Through the framework of reinforcement learning, we wish to produce a policy
function $\pi_\theta(a_t\mid s_t)$ that takes the observed binary spin state
$s_t\in\{-1,1\}^{\Nreps\times L}$ and produces an action $a_t$ corresponding to
the optimal change in the inverse temperature.

We now briefly introduce PPO \cite{Schulman2017}. First we define our policy $\pi_\theta(a_t\mid s_t)$ as the likelihood that the agent will take action $a_t$ while in state $s_t$; through training, the desire is that the best choice of action will become the most probable. To choose an action, this distribution can be sampled. We will use a neural network, parameterized by weights $\theta$ to represent the policy by assuming that $\pi_\theta (a_t \mid s_t)$ is a normal distribution and interpreting the output nodes of the neural network as the mean, $\mu$, and variance, $\sigma^2$.

We define a function $Q_{\pi_\theta}(s_t, a_t)$ as the expected future discounted reward if the agent takes action $a_t$ at time $t$ and then follows policy $\pi_\theta$ for the remainder of the episode.
We additionally define a value function $V_{\pi_\theta}(s_t)$ as the expected future discounted reward starting from state $s_t$ and following the current policy $\pi_\theta$ until the end of the episode. We introduce the concept of \emph{advantage}, $\hat{A}_t(s_t, a_t)$, as the difference between these two quantities. 
$Q_{\pi_\theta}$ and $V_{\pi_\theta}$ are not known and must be approximated.
We assume the features necessary to represent $\pi$ are generally similar to the features necessary to estimate the value function, and thus we can use the same neural network to predict the value function by merely having it output a third quantity.

$\hat{A}_t$ is effectively an estimate of how much better the agent did in choosing action $a_t$, compared to what was expected.
We construct the typical policy gradient cost function by coupling the advantage of a state--action pair with the probability of the action being taken,
\[ \loss{PG} = \mathbb{\hat{E}}_{t}\left[\log{\pi_\theta(a_t\mid s_t)}\hat{A}_t\right],\]
which we want to maximize by modifying the weights $\theta$ through the training process.  It is, however, more efficient to maximize the improvement ratio $r_t$ of the current policy over a policy from a previous iteration $\pi_{\theta_{\mathrm{old}}}$ \cite{schulman2015trust,kakade2002approximately}:

\[
\loss{TRPO} = \mathbb{\hat{E}}_t\left[\frac{\pi_\theta(a_t\mid s_t)}{\pi_{\theta_{\mathrm{old}}}(a_t\mid s_t)}\hat{A}_t\right]
\equiv
\mathbb{\hat{E}}_t\left[r_t(\theta)\hat{A}_t\right].
\]
Note, however, that maximizing this quantity can be trivially achieved by making the new policy drastically different from the old policy, which is not the desired behaviour.  The PPO algorithm \cite{Schulman2017} deals with this by clipping the improvement and taking the minimum
\[\loss{CLIP} = \mathbb{\hat{E}}_t\left[\min(r_t(\theta)\hat{A}_t, \mathrm{clip}(r_t(\theta), 1-\epsilon, 1+\epsilon)\hat{A}_t)\right].\]

To train the value function estimator, a squared error is used, that is,
\[ \loss{VF} = \mathbb{\hat{E}}_t[(V_{\pi_\theta}(s_t) - V_t^{\mathrm{targ}})^2], \]
and to encourage exploration, an entropic regularization functional $S$ is used. This all amounts to a three-term cost function

\[\loss{PPO} = \mathbb{\hat{E}}_t\left[\loss{CLIP} - c_1\loss{VF} + c_2 S[\pi_{\theta}](s_t)\right], \]
where $c_1$ and $c_2$ are hyperparameters.

\subsection{Policy network architecture}
The neural network is composed of two parts: a convolutional feature extractor,
and a recurrent network to capture the temporal characteristics of the problem.
The feature extractor comprises two parallel two-dimensional convolutional
layers. The first convolutional layer has $N_{k_r}$ kernels of size $1\times L$,
and aggregates along the replicas dimension, enabling the collection of
spin-wise statistics across the replicas.
The second convolutional layer has $N_{k_s}$ kernels of size $\Nreps \times 1$
and slides along the spin dimension, enabling the aggregation of replica-wise
statistics across the spins.
The outputs of these layers are flattened, concatenated, and fed into a dense
layer of size $N_d$ hidden nodes. This operates as a latent space encoding for
input to a recurrent neural network (a long short-term memory, or LSTM, module~\cite{Hochreiter1997}), used to capture the sequential nature of our
application. The latent output of the LSTM module is of size $N_L$. For
simplicity, we set $N_{k_r}=N_{k_s}=N_d=N_L=64$. All activation functions are
hyperbolic tangent ($\tanh$) activations. Since $a_t$ can assume a continuum of
real values, this task is referred to as having a continuous action space, and
thus standard practice is for the network to output two values corresponding to
the first and second moments of a normal distribution, which can be sampled to produce predictions.

\subsection{Reward}
At the core of \RL is the
concept of reward engineering, that is, developing a reward scheme to inject a notion
of success into the system. As we only care about reaching the ground state
by the end of a given episode, we use a sparse reward scheme, with a reward of zero
for every time step before the terminal step, and a reward equal to the negative
of the minimum energy as the reward for the terminal step, that is,
\begin{equation}
R_t = \begin{cases}
    0, & t < \Nsteps \\
    -\min\limits_{k}\mathcal{H}(\phi_k(s_t)), & t = \Nsteps
\end{cases},
\end{equation}
where $k\in[1,\Nreps]$, and
\[\phi_k(s_t) \in \{-1,1\}^{1\times L}\]
is an indexing function that returns the binary spin values for the $k$-th replica
of state $s_t$. \added{This reward function is agnostic to system size; as the system size increases, the correlation time will also increase, and additional sweeps may be required between actions, but the reward function remains applicable. Furthermore,} with this reward scheme, we encourage the agent to
arrive at the lowest possible energy by the time the episode terminates, without
regard to what it does in the interim.  In searching for the ground state, the
end justifies the means.

\subsection{Hyperparameters}

When optimizing the neural network, we use a PPO discount factor of
$\gamma=\numPPOGamma$, eight episodes between weight updates, a value function
coefficient of $c_1=0.5$, an entropy coefficient of $c_2=0.001$, a clip range of
$\epsilon=0.05$, a learning rate of $\alpha=\numLearningRate$, and a single
minibatch per update.  Each agent is trained over the course of $\numEps$
episodes (anneals), with $\Nsteps=\numNsteps$ steps per episode, and with
$\Nsweeps=\numNsweeps$ sweeps separating each observation. We used $\Nreps=\numNreps$
replicas for each observation.

\section{Evaluation}

Whereas the \RL policy can be made deterministic, meaning a given state always
produces the same action, the underlying Metropolis algorithm is stochastic; thus, we must statistically define the metric for success. We borrow this
evaluation scheme from Aramon~et~al.~\citep{Aramon2019}. Each \RL episode will either result in ``success'' or ``failure''.  Let us
define the ``time to solution'' as 
\begin{equation} 
T_{s}=\tau n_{99}\,,
\end{equation} 
that is, the number of episodes that must be run to be 99\% sure
the ground state has been observed at least one time ($n_{99}$), multiplied by
the time $\tau$ taken for one episode.  \added{As $\tau$ depends specifically on the hardware used, and the efficiency of software implementations, we will focus on $n_{99}$ alone as the metric we desire to minimize.}

Let us also define $X_i$ as the binary outcome of the $i$-th episode, with
$X_{i}=1~(0)$ if at least one (none) of the \Nreps replicas are observed to be
in the ground state at episode termination. The quantity $Y\equiv \sum_{i=1}^n
X_i$ is the number of successful episodes after a total of $n$ episodes, and
$p\equiv P(X_i = 1)$ denotes the probability that an anneal $i$ will be successful.
Thus the probability of exactly $k$ out of $n$ episodes succeeding is given by
the probability mass function of the binomial distribution

\begin{equation}\label{binomialdistribution}
 P(Y=k\mid n,p) = \begin{pmatrix}n\\k\end{pmatrix}  p^k(1-p)^{n-k}.
\end{equation}

To compute the time to solution, our quantity of interest is the number of
episodes $n_{99}$ where $P=0.99$, that is,
\[ P(Y\ge 1 \mid n_{99}, p) = 0.99.\]
From this and (\ref{binomialdistribution}), it can be shown that
\[n_{99} = \frac{\log{(1-0.99)}}{\log{(1-p)}}.\]

\omit{
This statement is equivalent to its complement
\[1-P(Y=0\mid n_{99}, p) = 0.99, \]
and we can solve for $n_{99}$ through substitution into the binomial
distribution (\Eqref{binomialdistribution}).

\begin{align*}
    0.99 &= 1-P(Y=0\mid n_{99}, p)\\
    &=1-\begin{pmatrix}n_{99}\\0\end{pmatrix}(1-p)^{n_{99}-0} \\
    &=1-(1-p)^{n_{99}} \\
n_{99} &= \frac{\log{(1-0.99)}}{\log{(1-p)}}.
\end{align*}
}

In the work of Aramon~et~al.~\cite{Aramon2019}, $p$ is estimated using Bayesian inference due to their
large system sizes sometimes resulting in zero successes, precluding the direct
calculation of $p$.  In our case, to evaluate a policy, we perform \numTestRuns
runs for each of \numTestInstances instances and compute $p$ directly from the
ratio of successful to total episodes, that is, $p = \bar X$.

\section{Hamiltonians}
\label{sec:Hamiltonians}

We present an analysis of two classes of Hamiltonians. The first, which we call
the weak-strong clusters model (WSC; see \figref[a]{fig:hamiltonians}), is an $L=16$
bipartite graph with two fully connected clusters, inspired by the ``Chimera''
structure used in D-Wave Systems' quantum annealing hardware \cite{Bunyk2014}. In our
case, one cluster is negatively biased with $h_i=-0.44$ and the other
positively biased with $h_i=1.0$. All couplings are ferromagnetic and have unit magnitude.
This results in an energy landscape with a deep local minimum where both
clusters are aligned to their respective biases, but a slightly lower global
minimum when the two clusters are aligned together, sacrificing the benefit of
bias-alignment for the satisfaction of the intercluster couplings.  For all \WSC
runs, the spins of the lattice are initialized in the local minimum.

The second class of Hamiltonians are nearest-neighbour square spin glasses (SG; see~\figref[b]{fig:hamiltonians}). Couplings are periodic (i.e., the model is defined
on a torus), and drawn from a normal distribution with standard deviation 1.0.
All biases are zero. Hamiltonian instances are generated as needed during
training. To evaluate our method and compare against classic SA, we must have a testing set of instances for which we know the true
ground state. For each lattice size investigated
($\sqrt{L} = [4, 6, 8, 10, 12, 14, 16]$) we generate $N_\mathrm{test}=100$ unique
instances and obtain the true ground state energy for each instance using the
branch-and-cut method \cite{Liers2005} through the Spin Glass Server
\cite{Junger}.

\section{Results}

\begin{figure*}
    \centering
    \includegraphics[width=\textwidth]{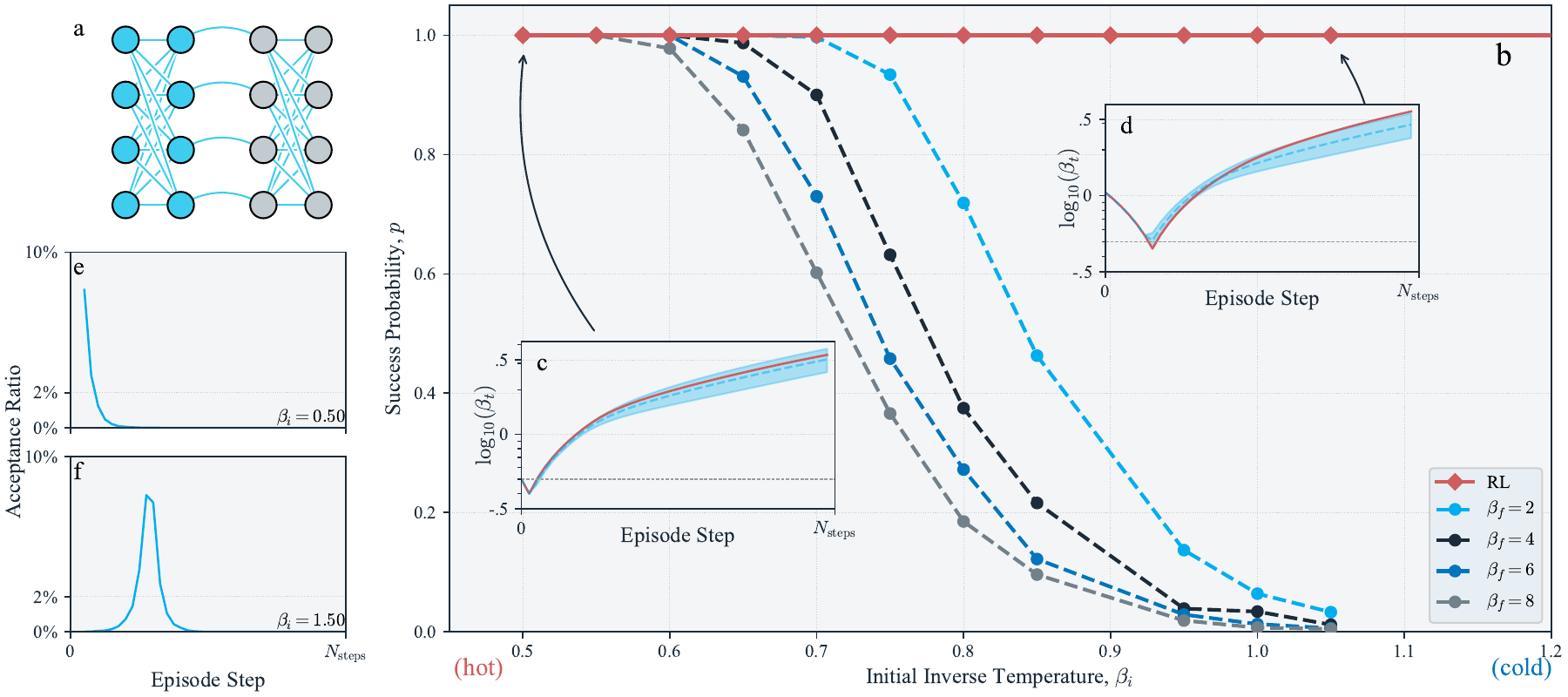}
    \caption{
    An \RL policy learns to anneal the \WSC model (shown in (a)).  (b) We plot the
    performance of various classic SA schedules,
    cooling linearly from $\beta_i$ to $\beta_f$, as well as the performance of the \RL
    policy for a variety of starting temperatures.  When the initial inverse
    temperature is sufficiently small, both the \RL and classic SA algorithms achieve
    100\% success (i.e., every episode reaches the ground state).  When the
    system is initialized with a large $\beta_i$, there is insufficient thermal
    energy for classic SA to overcome the energy barrier and reach the ground state, and
    consequently a very low success probability. A single \RL policy achieves
    almost perfect success across all initial temperatures. In (c) and (d), we plot
    the \RL inverse temperature schedule in red for  episodes initialized with  respective low and high inverse temperatures. In blue, we show the average \RL policy
    for the specific starting temperature.  The \RL
    algorithm can identify a cold initialization from observation, and increases the temperature
    before then decreasing it (as shown in (d)). In (e) and (f), we plot the
    Metropolis acceptance ratio for two episodes, initialized at two extreme
    temperatures (e) low $\beta_i$, and (f) high $\beta_i$. In this work, we use $\Nsteps=40$ episode steps.}
    \label{fig:wscresults}
\end{figure*}

\subsection{Weak-strong clusters model}

We demonstrate the use of \RL on the \WSC model shown in
\figref[a]{fig:hamiltonians}. RL is able to learn a simple temperature schedule
that anneals the \WSC model to the ground state in 100\% of episodes, regardless
of the temperature in which the system is initialized.  In
\figref[b]{fig:wscresults}, we compare the RL policy to  classic SA schedules with several constant cooling rates.

When the system is initially hot (small $\beta$), both \RL and classic SA are capable of
reaching the ground state with 100\% success as there exists sufficient thermal
energy to escape the local minimum.  In \figref[c]{fig:wscresults}, we plot an
example schedule.  The \RL policy (red) increases the temperature slightly at
first, but then begins to cool the system almost immediately.  An abrupt
decrease in the Metropolis acceptance rate is observed
(\figref[e]{fig:wscresults}).  The blue dashed line in
\figref[c]{fig:wscresults} represents the average schedule of the \RL policy over 100
independent anneals.  The standard deviation is shaded.  It is apparent that the
schedule is quite consistent between runs at a given starting temperature, with
some slight variation in the rate of cooling.

When the system is initially cold (large $\beta$), there exists insufficient
thermal energy to overcome the energy barrier between the local and global minima, and SA fails with a constant cooling rate. The \RL policy, however, is
able to identify, through observation, that the temperature is too low and can
rapidly decrease $\beta$ initially, heating the system to provide sufficient
thermal energy to avoid the local minimum.  In \figref[f]{fig:wscresults}, we see
an increase in the Metropolis acceptance ratio, followed by a decrease,
qualitatively consistent with the human-devised heuristic schedules that have been
traditionally suggested
\cite{VanLaarhoven1987,Kirkpatrick1983a,Kirkpatrick1984}. In
\figref[d]{fig:wscresults}, we plot an example schedule.  Initially, the \RL
algorithm increases the temperature to provide thermal energy to escape the
minimum, then begins the process of cooling.  Similar to
\figref[c]{fig:wscresults}, the broadness of the variance of the policies is
greatest in the cooling phase, with some instances being cooled more rapidly
than others. The \RL agent does not have access to the current temperature
directly, and bases its policy solely on the spins.  The orthogonal unit-width
convolutions provide a mechanism for statistics over spins and replicas, and the
LSTM module provides a mechanism to capture the time-dependent dynamics of the
system.

\begin{figure*}
    \centering
    \includegraphics[width=\textwidth]{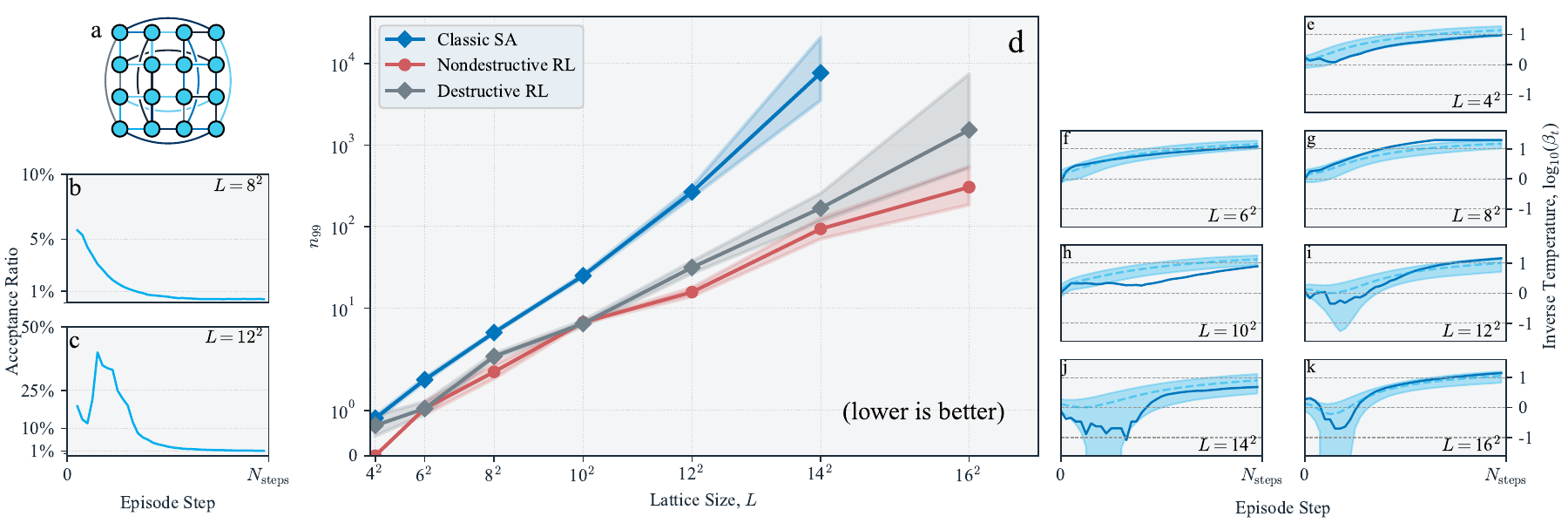}
    \caption{
    An RL policy learns to anneal spin-glass models. An example ($L=4^2$)
    lattice is shown in (a). In (b) and (c), we plot the acceptance ratios over
    time for three episodes for each of the $L=8^2$ and $L=16^2$ lattices. In
    (d), we compare the scaling of the \RL policy with respect to system size and
    compare it to classic SA.  We plot the
    $n_{99}$ value (the number of anneals required to be 99\% certain of
    observing the ground state\added{; in the case of 100\% success, $n_{99}$ is undefined and plotted as zero}) as a function of system size for both the
    \RL and the best linear simulated annealing schedule we
    observed. \added{The 95\% confidence interval is shown as a shaded region.} For all system sizes investigated, the learned \RL policy is able
    to reach the ground state in significantly fewer runs. Additionally, we plot
    the destructive observation results, which also outperform the linear schedules.  We note that the destructive observation
    requires far more Monte Carlo steps per episode to simulate the destructive
    measurements; this plot should not be interpreted as a comparison of run
    time with regard to the destructive observation result. In (e) through (k),
    we plot an example inverse temperature schedule as a solid line, as well as
    the average inverse temperature schedule (for all testing episodes) as a
    dashed line, with the shaded region denoting the standard deviation.  \added{In this work, we use $\Nsteps=40$ episode steps.}}
    \label{fig:spinglassresults}
\end{figure*}

\subsection{Spin-glass model}

We now investigate the performance of the \RL algorithm in learning a general
policy for an entire class of Hamiltonians, investigating whether the \RL
algorithm can learn to generalize its learning to accommodate a theoretically
infinite set of Hamiltonians of a specific class. Furthermore, we investigate
how \RL performs with various lattice sizes, and compare the trained \RL model
to a linear (with respect to $\beta$) classic SA schedule such as the ones used \cite{stander1994temperature,Heim215}. The results of
this investigation are shown in \figref{fig:spinglassresults}.

In all cases, the \RL schedule obtains a better (lower)
$n_{99}$ value, meaning far fewer episodes are required for us to be confident
that the ground state has been observed.  Furthermore, the $n_{99}$ value
exhibits much better scaling with respect to the system size (i.e., the number of
optimization variables). In \figref[e--k]{fig:spinglassresults}, we plot some of
the schedules that the \RL algorithm produces. In many cases, we see initial
heating, followed by cooling, although in the case of the larger models (\figref[i--k]{fig:spinglassresults}) we
see much more complex, but still successful, behaviour. In all cases, the
variance of the policies with respect to time (shown as the shaded regions in
\figref[e--k]{fig:spinglassresults}), indicate the agent is using information
from the provided state to make decisions, and not just basing its decisions on
the elapsed time using the internal state of the LSTM module. If schedules were
based purely on some internal representation of time, there would be no variance
between episodes.

\added{
\begin{figure*}
    \centering
    \includegraphics[width=\textwidth]{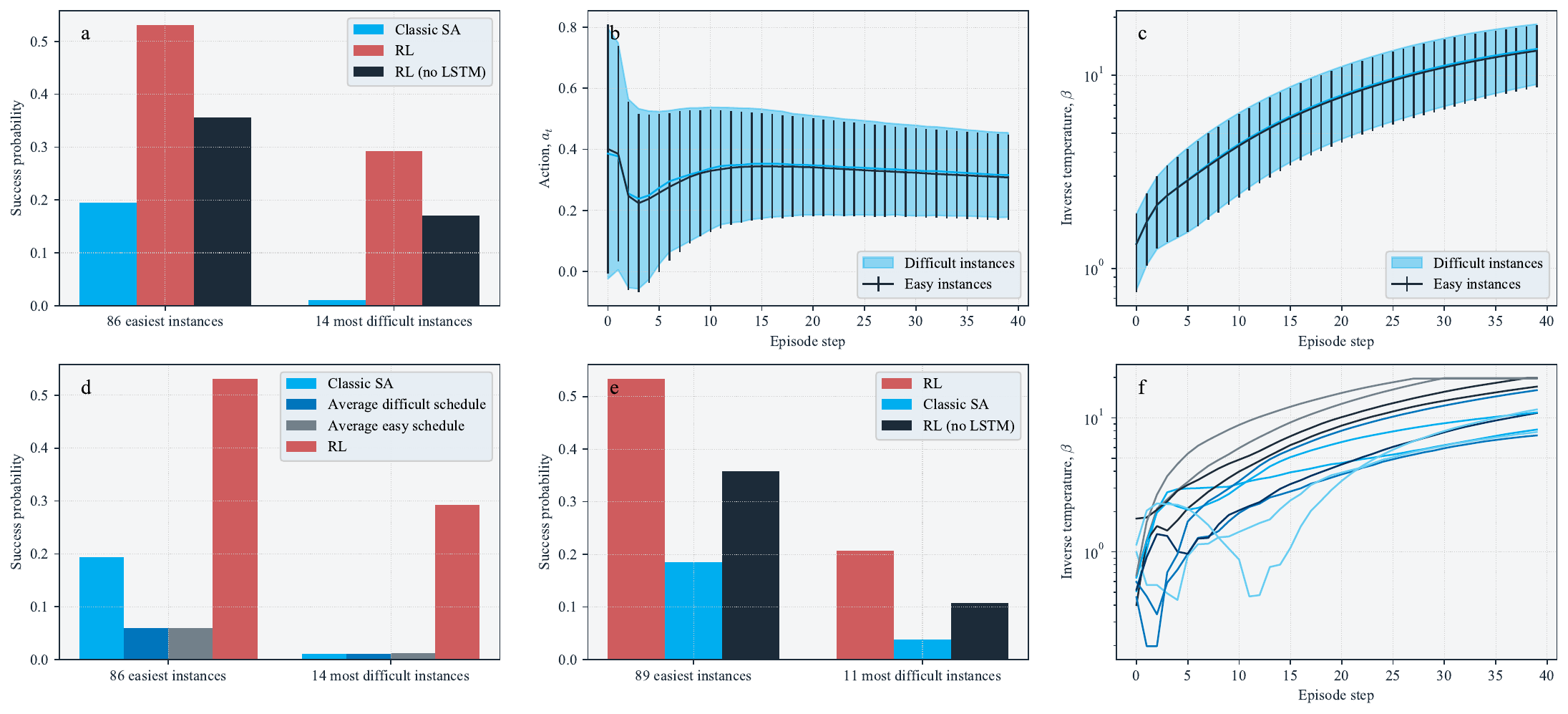}
    \caption{
    \added{
    We separate the $10\times10$ spin glass instances in the test set into two subsets (easy and difficult), depending on the success of classic SA in finding their ground states.  In (a) we plot the performance of three different temperature scheduling approaches on these subsets.  RL exhibits superior performance over classic SA in both subsets; however, it demonstrates dramatic superiority in the case of the difficult instances.  RL without an LSTM module still performs better than classic SA; it can still dynamically modify the schedule and is not constrained to a constant temperature change at each step, so is more akin to a traditional heuristic temperature scheduling approach.  In (b) and (c) we plot the average RL actions and schedule, respectively, for both the difficult and easy instance subsets. The standard deviation of the policies are plotted with error bars (easy instances) and shaded regions (difficult instances).  The average difficult policy is very similar to the average easy policy, both having a large standard deviation, suggesting a high degree of specificity of the policy to the given episode.  We can see this in (f), where we plot several successful schedules; each schedule is quite different from the others, but each results in a successful episode.  In (d), we show the performance when we apply the average actions presented in (b) as a static policy.  The average policies perform even more poorly than classic SA. This is further evidence that the RL agent's ability to observe the system is crucial to its high performance. One might object to the method used to split the instances into the difficult and easy subsets; we have explicitly chosen to split the subsets at a boundary that makes classic SA perform poorly on the difficult instances. Therefore in (e), we consider a difficult (easy) instance as one that the RL agent performs poorly (well) on, and the story remains unchanged.}
    }
    
    \label{fig:easy_diff}
\end{figure*}

\subsection{Comparing easy and difficult instances}

The learned strategy of the RL agent is relatively simple in concept: increase the temperature to a sufficiently high value and then use the remaining time to cool the system as seen in the average policies in \figref[e--k]{fig:spinglassresults}.  In this section we demonstrate the degree to which the performance improvement can be attributed to the ability of the RL agent to base its decisions upon the various dynamics in the system.

We divide the instances in the $10\times 10$ test set into two 
subsets, which we label ``easy'' and ``difficult'', based on the success 
of the classic SA baseline. This results in 14 difficult instances in which classic 
SA succeeds in only 3\% of anneals, and 86 easy instances in which classic SA 
succeeds in more than 3\% of anneals.

We compare three temperature scheduling methods on 100 episodes of each instance in both of these subsets: 
i) classic (linear) SA; 
ii) the RL agent; and 
iii) an RL agent (not yet discussed) that does not include a recurrent LSTM module.  
As shown in \figref[a]{fig:easy_diff}, 
linearly scheduled classic SA solves the easy instances in 19\% 
of anneals, whereas the reinforcement learning agent manages to solve the same instances with a 
53\% success probability.
With the difficult instances, the difference is more extreme; classic SA manages only 
1\% success, whereas 
RL performs substantially better with 29\% 
success.

A variant of the agent without an LSTM module performs more poorly, but still better than classic SA.  This agent
is simply provided with a floating point representation of the episode step concatenated to the state vector,
but without a recurrent network, it has no mechanism to capture the time dependence (history) of the problem. It therefore
can only use the current observation in making decisions, and evidently does so more poorly than the agent with
access to an LSTM module. For our formulation of the environment, an LSTM module is theoretically important to achieve a well-defined Markov decision process.

In \figref[b]{fig:easy_diff}, we plot the average action taken, and in  \figref[c]{fig:easy_diff}, we plot the average inverse temperature of the
system at each step in the test episodes driven by the RL agent, averaged over the easy and difficult
instances separately.
There is no notable difference in the average schedules of the two subsets. This fact, combined with the considerable magnitude of the 
standard deviation (plotted as a filled region for difficult instances and vertical bars for easy ones) suggests
that the RL agent is adaptive to the specific instantiation of each Hamiltonian. Some of these dynamics can be seen in the successful schedules randomly selected for plotting in \figref[f]{fig:easy_diff}.

We then take the average schedules plotted in \figref[b--c]{fig:easy_diff} and use them as if they were  RL-designed
general heuristic schedules, removing the necessity to conduct observations during the training procedure. Both the difficult
and easy average schedules perform very poorly on both the difficult and easy subsets, succeeding in less than 10\% of episodes. This is strong evidence of the 
specificity of the RL agent's actions to the particular dynamics of each episode and refutes the hypothesis that a single, average policy, even if trained by RL, is a good case for generic instances.

We repeat the previous analysis with subsets based on the performance of the RL agent, arriving at identical conclusions (\figref[e]{fig:easy_diff}).

}

\subsection{Destructive observation}

A key element of the nature of quantum systems is the collapse of the wavefunction when a measurement of the quantum state is made. When dealing with quantum systems, one must make control decisions based on
quantum states that have evolved through an identical policy but have never
before been measured. We model this restriction on quantum measurements by
allowing any replica observed in the anneal to be consumed as training data for
the \RL algorithm only once. We simulate this behaviour by keeping track of the
policy decisions (the changes in inverse temperature) in an action buffer as we
play through each episode.  When a set of $\Nreps$ replicas are measured, they
are consumed and the system is \added{reset to a new set of initial conditions, as if it was a new episode}. The actions held in the buffer are
replayed on the new replicas.

In this situation, the agent cannot base its decision on any replica-specific
temporal correlations between given measurements; this should not be a problem
early in each episode, as the correlation time scale of a hot system is very
short, and the system, even under nondestructive observation, would have evolved
sufficiently, in the time window between steps, to be uncorrelated.  However, as
the system cools, the correlation time scale increases exponentially, and
destructive observation prevents the agent from relying on  temporal correlations of
any given replica.

We evaluate an agent trained in this ``quantum-inspired'' way and plot its
performance alongside the nondestructive (i.e., classical) case in
\figref[d]{fig:spinglassresults}. In the case of destructive observation, the agent performs
marginally less well than the nondestructive case, but still performs better than
 SA in most cases. As it is a more complicated task to make
observations when the system is temporally uncorrelated, it is understandable
that the performance would be inferior to the nondestructive case.  Nonetheless,
\RL is capable of outperforming SA in both the destructive and
nondestructive cases.

The relative performance in terms of computational demand between 
destructive observation and SA alludes to an important future
direction in the field of \RL, especially when applied to
physical systems where observation is destructive, costly, and altogether
difficult. With destructive observations, \Nsteps systems must be initialized
and then evolved together under the same policy.  Each copy is consumed
one by one, as observations are required for decision making, thus incurring an
unavoidable $\Nsteps^2/2$ penalty in the destructive case.  In this sense, it is
difficult to consider \RL to be superior; prescheduled SA simply does not require observation.  However, if the choice to
observe were to be incorporated into the action set of the \RL algorithm, the agent would choose when observation would be necessary.

For example, in the systems presented in this work, the correlation time of the
observations is initially small; the temperatures are high, and frequent
observations are required to guide the system through phase space.  As the
system cools, however, the correlation time grows exponentially, and the observations become much more similar to each previous observation; in this case, it would be
beneficial to forgo some expensive observations, as the system would not be
evolving substantially.  With such a scheme, RL stands a better chance at achieving greater performance.

\subsection{Policy analysis}
\label{sec:policyanalysis}

\begin{figure*}
    \centering
    \includegraphics[width=\textwidth]{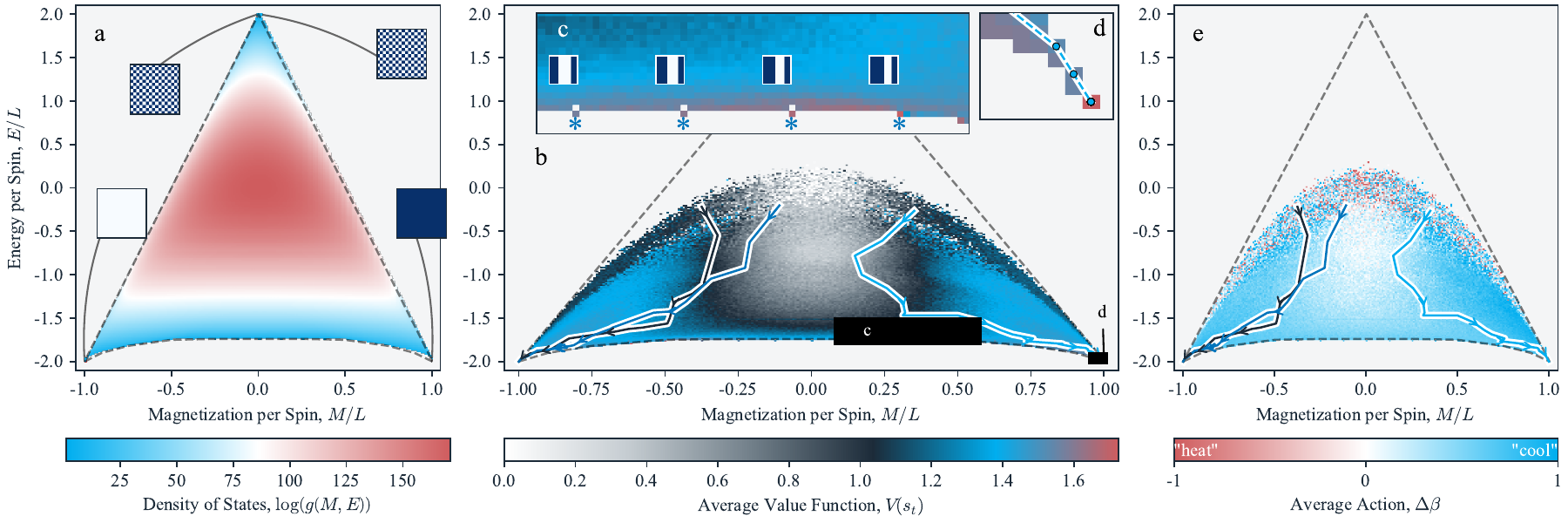}
    \caption{
    We train an agent on a special case of the spin-glass Hamiltonians:
    the $16\times 16$ ferromagnetic Ising model where all couplings $J_{ij}=1$.
    (a) We plot the density of states $\log(g(M, E))$ for the $16\times 16$
    Ising model in the phase space of energy and magnetization, sampled
    numerically using the Wang--Landau algorithm \cite{Wang2001}, and indicate
    four of the novel high- and low-energy spin configurations on a grid. (b)
    For the trained model, we plot the average of the learned value function
    $V(s_t)$ for each possible energy--magnetization pair. Additionally, we plot
    the trajectories of the first replica for three episodes of annealing to
    demonstrate the path through phase space the algorithm learns to take. In
    (c) and (d), we enlarge two high-value regions of interest. In (e), we plot
    the average action taken at each point in phase space, as well as the same
    two trajectories plotted in (b).}
    \label{fig:valueplot}
\end{figure*}

To glean some understanding into what the \RL agent is learning, we train an
additional model on a well-understood Hamiltonian, the ferromagnetic Ising model
of size $16\times 16$. In this case, the temperatures are initialized randomly
(as in the \WSC model). This model is the extreme case of a spin glass, with all
$J_{ij}=1$. In \figref[a]{fig:valueplot}, we display the density of states $g(M,
E)$ of the Ising model, plotted in phase space, with axes of magnetization per
spin ($M/L$) and energy per spin ($E/L$).  The density of states is greatest
in the high-entropy $M=E=0$ region, and lowest in the low-entropy ``corners''.
We show the spin configurations at the three corners (``checkerboard'', spin-up, and spin-down) for clarity. The density of states is obtained numerically using
Wang--Landau sampling \cite{Wang2001}. Magnetization and energy combinations
outside of the dashed ``triangle'' are impossible.

In \figref[b]{fig:valueplot}, we plot a histogram of the average value function
$V(s_t)$ on the phase plane, as well as three trajectories. Note that since each
observation $s_t$ is composed of $\Nreps$ replicas, we count each observation
as $\Nreps$ separate points on the phase plot when computing the histogram, each
with an identical contribution of $V(s_t)$ to the average. As expected, the learned
value function trends higher toward the two global energy minima.  The lowest values are present in the initialization region (the
high-energy band along the top). We expand two regions of interest in
\figref[c--d]{fig:valueplot}. In \figref[d]{fig:valueplot}, we can see that the global minimum is
assigned the highest value; this is justifiable in that if the agent
reaches this point, it is likely to remain here and reap a high reward so long
as the agent keeps the temperature low for the remainder of the episode.

In \figref[c]{fig:valueplot}, we identify four noteworthy energy--magnetization
combinations, using asterisks. These four energy--magnetization combinations have
identical energies, with increasing magnetization, and correspond to banded spin
structures of decreasing width (four example spin configurations are shown).
The agent learns to assign a higher value to the higher-magnetization
structures, even though the energy, which is the true measure of ``success'', is
identical.  This is because the higher-magnetization bands are closer to the
right-most global minimum in action space, that is, the agent can traverse from the small-band configuration to the ground state in fewer spin flips than if traversing from the
wide-band configurations.

In \figref[e]{fig:valueplot}, we plot a histogram of the average action taken at
each point in phase space. The upper high-energy band exhibits more randomness
in the actions chosen, as this is the region in which the system lands upon
initialization. When initialized, the temperature is at a randomly drawn value,
and sometimes the agent must first heat the system to escape a local minimum
before then cooling, and thus the first action is, on average, of very low
magnitude.  As the agent progresses toward the minimum, the agent becomes
more aggressive in cooling the system, thereby thermally trapping itself in lower energy states.

\added{
\subsection{Scaling and time to solution}
\figref[d]{fig:spinglassresults} indicates that both nondestructive and destructive \RL perform substantially better not only in absolute terms, but also in terms of scaling. It is important to note that we have specifically chosen a neural network architecture (convolutional) that scales linearly with system size, and have trained each model for the same number of episodes, each consisting of the same number of sweeps. The computation time for each sweep scales linearly with the system size, and thus the training time of our \RL models scales linearly with system size.  Using RL does indeed impose an additional inference cost, as the observation must be processed by the neural network; on the $L=10^2$ system, inference takes one-third the amount of time as does each episode step. However, this cost has not been optimized, and could significantly be lowered through optimization of the neural network inference or even by offloading the policy network onto specialized hardware designed for inference.
}

\section{Conclusion}

In this work, we show that reinforcement learning is a viable method for learning
dynamic control schemes for the task of simulated annealing (SA).  We show that, on a
simple spin model, a reinforcement learning (RL) agent is capable of devising a
temperature control scheme that can consistently escape a local minimum, and then
anneal to the ground state.  It arrives at a policy that generalizes to a range
of initialization temperatures; in all cases, it learns to cool the system. However,
if the initial temperature is too low, the \RL agent learns to first increase
the temperature to provide sufficient thermal energy to escape the local minimum. It achieves this without being provided explicit knowledge of the temperature.

We then demonstrate that the \RL agent is capable of learning a policy that can
generalize to an entire class of Hamiltonians, and that the problem need not be
restricted to a single set of couplings. By training multiple \RL agents on
increasing numbers of variables (increasing lattice sizes), we investigate the
scaling of the \RL algorithm and find that it outperforms a
classic SA schedule both in absolute terms and in terms of its
scaling.

\added{Our technique is not limited to the system sizes we present in this work; larger system sizes are also within its reach.  At some point, as the size of the system increases, correlation times in the underlying Metropolis--Hastings simulation become larger than the intervals between  observations,  and  the  number  of  sweeps  must be increased.  Additionally, we have specifically chosen a neural network architecture that scales linearly with system size (convolutional neural networks) as opposed to traditional multi-layer perceptron networks that scale exponentially. In fact, the entire procedure scales at most polynomially with system size.  }

We analyze the value function that the agent learns and see that it attributes
an intuitive representation of value to specific regions of phase
space.

We discuss the nature of \RL in the physical sciences,
specifically in situations where observing systems is destructive (``destructive
observation'') or costly (e.g., performing quantum computations where observations collapse
the wavefunction, or conducting chemical analysis techniques that destroy a sample material).
We demonstrate that our implementation of \RL is capable of performing well in a
destructive observation situation, albeit inefficiently.  We propose that the
future of physical \RL (i.e., \RL in the physical sciences) will be one of
``controlled observation'', where the algorithm can choose when an observation
is necessary, minimizing the inherent costs incurred when observations are
expensive, slow, or difficult.

\section{Correspondence}
Requests for materials can be made to any of the authors. The code and data is freely available at the source provided below. 

\section{Acknowledgements}

I. T. acknowledges support from NSERC. K. M. acknowledges support from \mbox{Mitacs}. The authors would like to thank Bruce Krayenhoff for valuable discussions in the early stages of the project, and would like to thank Marko Bucyk for reviewing and editing the manuscript.

\section{Statement of Contributions}
All authors contributed to the ideation and design of the research. K. M. developed  and ran the computational experiments,
and wrote the initial draft of the the  manuscript.  P. R. and I. T. jointly supervised this work and revised the manuscript.

\section{Data availability}
The test data sets necessary to reproduce these findings are available at \href{https://doi.org/10.5281/zenodo.3897413}{https://doi.org/10.5281/zenodo.3897413}.

\section{Code availability}

The code necessary to reproduce these findings is available at \href{https://doi.org/10.5281/zenodo.3897413}{https://doi.org/10.5281/zenodo.3897413}.

\bibliographystyle{bst/naturemag}
\bibliography{bib}

\end{document}